\documentclass[aps,prb,preprint,superscriptaddress,showpacs]{revtex4-1}

\usepackage{textcomp}
\usepackage{amsmath}
\usepackage{amssymb}
\usepackage{bm}
\usepackage{graphicx}
\usepackage{xcolor}
\usepackage{dcolumn} %align numbers by decimal point; note that siunitx does not seem to work with ruledtabular and revtex4-1
\usepackage[caption=false]{subfig} % apparently RevTeX4 is not compatible with {caption}, which is required by {subcaption}
\usepackage{footnote}
\usepackage{hyperref}
\usepackage{enumerate}
\usepackage{epstopdf}
\usepackage{chngcntr}

\newcommand{\kvec}{$\mathbf{k}$}
\newcommand{\qvec}{$\mathbf{q}$}

\newcommand{\qe}{{\sc Quantum ESPRESSO}}
\newcommand{\qeabbrev}{{\sc QE}}

\newcolumntype{d}[1]{D{.}{.}{#1}}
\newcolumntype{e}[1]{D{.}{}{#1}}

\begin{document}

% =========================
% TITLE
% =========================
\title{Metallic Hydrogen: A Liquid Superconductor?}

% =========================
% AUTHORS & AFFILIATIONS
% =========================
\author{Craig M.\ Tenney}
%\email[]{craig.tenney@wsu.edu}
\affiliation{Department of Physics and Astronomy, Washington State University, Pullman, Washington 99164, USA}

\author{Zachary F.\ Croft}
\affiliation{Department of Physics and Astronomy, Washington State University, Pullman, Washington 99164, USA}

\author{Jeffrey M.\ McMahon}
\email[]{jeffrey.mcmahon@wsu.edu}
\affiliation{Department of Physics and Astronomy, Washington State University, Pullman, Washington 99164, USA}

% =========================
% DATE
% =========================
\date{\today}

% =========================
% ABSTRACT
% =========================

\begin{abstract}
Metallic hydrogen is expected to exhibit remarkable physics. Of particular interest in this work is the possibility of high-temperature superconductivity. Comparing calculations of the superconducting critical temperatures of the solid phase to melting temperatures over a range of pressures leads to an interesting question: Will the solid, in a superconducting state, melt to a liquid that remains a superconductor? In this work, the possibility of liquid superconductivity in metallic hydrogen is investigated. This is done by first-principles simulations, and using the results of these to solve the Eliashberg equations. These are carried out over the pressure (and temperature) conditions where molecular dissociation is expected to first occur in the solid phase. Over the pressure range $386.8(4)$--$783.7(4)$ GPa, $T_c$ increases from $308(6)$ to $372(2)$ K with a maximum uncertainty of $10$ K; it then decreases to $356(2)$ K at $883.7(3)$ GPa. Comparisons to the solid phase show that the critical temperature is not significantly changed between the two phases, though the physics behind their superconductivity is different. Careful comparisons of these values to recent results in the context of the hydrogen phase diagram show that they are higher than the melting temperatures and that the solid will melt to liquid atomic hydrogen. The results of this work (in this context) therefore suggest that liquid atomic hydrogen will indeed exist in a superconducting state. They also provide the pressure and temperature conditions over which to look for it.
\end{abstract}

% =========================
% PACS & KEYWORDS
% =========================


%\keywords{}

% =========================
% MAKETITLE
% =========================
% must follow title, authors, abstract, \pacs, and \keywords

\maketitle

% =========================
% BODY OF PAPER
% =========================
% use proper section commands
% references should be done using the \cite, \ref, and \label commands

\section{Introduction}
\label{sec:intro}

\begin{figure}[h]
    \centering
    \includegraphics[width = 7.5cm]{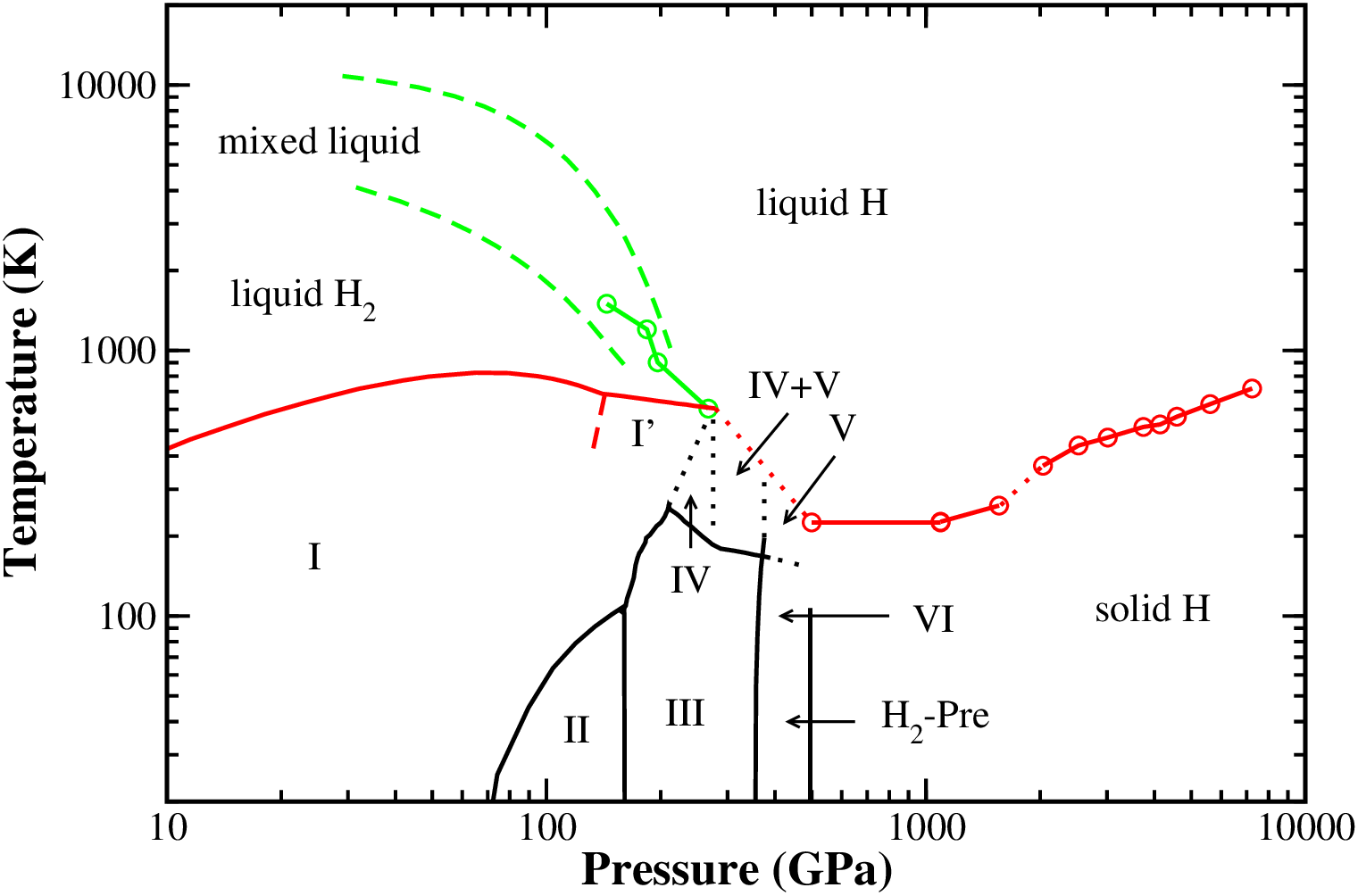}
    \caption{Phase diagram of hydrogen. Data combined from experimental and computational studies \cite{PhysRevLett.119.065301, Eremets2011, eremets2016low, Dalladay-Simpson2016, PhysRevB.100.184112, PhysRevLett.114.105305, Diaseaal1579, eaao5843, PhysRevLett.119.075302, doi:10.1021/jp403885h, PhysRevB.92.104103, Geng2016, https://doi.org/10.1002/advs.201901668, Pierleoni4953}.} 
    \label{fig:phase_diag}
\end{figure}

Hydrogen is the most abundant element in the Universe, comprising roughly $74$\% of all baryonic matter. Despite its chemical simplicity (a single proton and electron), its behavior over a wide range of thermodynamic conditions is remarkably complex. This can be seen, for example, in the hydrogen phase diagram, as shown in Fig.\ \ref{fig:phase_diag}.
(Many of the specific details of this diagram are not necessary to consider; rather, this diagram will be used for reference during discussion herein.) With this complexity comes remarkably rich physics. Of interest in this work are the properties of dense hydrogen (in particular, as caused by high pressures). These have been generally reviewed in Ref.\ \onlinecite{RevModPhys.84.1607}, and those of particular interest are done so below.

In 1935, Wigner and Huntington predicted \cite{doi:10.1063/1.1749590} that sufficient pressure would dissociate hydrogen molecules, and that any Bravais lattice of such atoms would be metallic. The pressures required to dissociate hydrogen molecules are expected to be significant [$447(3)$ GPa according to computation \cite{PhysRevLett.114.105305}, consistent with experiment that shows \cite{Eremets2019} at least above $440$ GPa, possibly near $495$ GPa \cite{Diaseaal1579, eaao5843}]. This transition can be seen by the vertical line in Fig.\ \ref{fig:phase_diag}. 

% NOTE: The physics in (iii) are typical features of a high-density system.
%
% NOTE: The trend in T_c with pressure is for the most stable atomic solid phase.
%
In 1968, Ashcroft predicted \cite{PhysRevLett.21.1748} another type of transition, a metallic-to-superconducting one. Within the framework of Bardeen--Cooper--Schrieffer (BCS) theory \cite{PhysRev.106.162, PhysRev.108.1175}, three arguments were made to support this prediction: (i) the ions in the system are single protons, and their small mass causes the vibrational energy scale of the phonons to be remarkably high (this generally enters as a prefactor in approximate expressions for the superconducting critical temperature $T_c$); (ii) since the electron--ion interaction is due to the bare Coulomb attraction, the electron--phonon coupling should be strong; and (iii) at the high pressures at and above metallization, the electronic density of states $N(0)$ at the Fermi surface should be large and the Coulomb repulsion between electrons relatively low. Recent calculations \cite{PhysRevB.93.174308} near the (solid) molecular-to-atomic transition give a value of $T_c = 300$ K at $500$ GPa. Calculations \cite{PhysRevB.84.144515, PhysRevB.85.219902} further show an increase with increasing pressure with a maximum near $700$ GPa.

Recent calculations of the melting line of the atomic solid \cite{doi:10.1021/jp403885h, PhysRevB.92.104103} (discussed above) show that it melts around $200$--$250$ K near and just above the pressure of molecular dissociation. This range is indicated on the phase diagram in Fig.\ \ref{fig:phase_diag} as a horizontal line at $225$ K. (A more thorough and specific discussion of these results is reserved for later, so that they can be considered in the context of those presented herein.) If these results are considered with those above, it can be seen that the $T_c$ values are even higher. This leads to an interesting question: Will the solid, in a superconducting state, melt to a liquid that remains a superconductor?

% NOTE: Compressibility data between solids and liquids can be found in a constants handbook.
%
% NOTE: Edwards, et al. further discusses the possibility of non-BCS liquid superconductors. 
%
In 1981, Jaffe and Ashcroft gave \cite{PhysRevB.23.6176} two reasons for why liquid superconductors could theoretically exist: (i) compressibility of metals changes very little during melting, and so high-frequency, longitudinal phonons should be similar in both the solid and liquid phases; and (ii) the existence of amorphous superconductors indicates that disorder does not inhibit superconductivity. For all classical metals, however, BCS liquid superconductors seem almost implausible. Arguments \cite{edwards2006possibility} based on comparisons of $T_c$ to melting temperatures have been used to show this. This is not the case though for quantum metals --- systems with large quantum zero-point motion of the ions that may strongly depress the (classical) melting melting temperature. This would be the case, for example, for some light elements. The resultant liquid must also be metallic though, meaning that dense hydrogen might be the \emph{only} such system where this would be plausible. This problem has received relatively little attention, however. And now that experiments (such as those discussed above, and will also be discussed below) are now capable of or are approaching the relevant thermodynamic conditions makes it a timely problem to consider.       

In this Article, the possibility of superconductivity in liquid atomic hydrogen is investigated by first-principles simulations, and using the results of these to solve the Eliashberg equations \cite{Eliashberg_JETP-1960}. This is done over thermodynamic conditions where molecular dissociation is expected to first occur and where superconductivity in the liquid would be most likely (if it occurs at all). This would be approximately $350$--$850$ GPa and less than $500$ K. 
% phase diagram: These conditions are indicated as those below a dotted line added Fig.\ \ref{fig:phase_diag}.

This Article is outlined as follows. In the next section (Section \ref{sec:methods}), a method is carefully considered to calculate properties associated with superconductivity for a liquid phase. Section \ref{sec:results} presents and discusses the results from the application of this method to liquid atomic hydrogen. Section \ref{sec:discussion} concludes, by discussing these results in the context of the hydrogen phase diagram and current experimental techniques. 
% SI: A Supplementary Information (SI) accompanies this work.

% NOTE: CMT discusses convergence tests (k points and cutoffs) on a bcc lattice at 550 GPa. More specific details are not provided.
%
\section{Methods}
\label{sec:methods}

Calculating superconductivity properties is relatively straightforward for a solid, by using methods developed to study lattice dynamics. Things become more involved when considering a liquid, due to dynamical motion. Under some careful considerations and reasonable approximations though, it is possible to leverage and apply the aforementioned methods to the system of interest.

The method considered in this work consists of the following steps:
\begin{enumerate}
	\item Optimization of starting solid structures, to determine constant volumes representative of desired pressures.
	\item Molecular dynamics simulations, to melt the solids and then simulate the liquid at constant temperature.
	\item Determination of statistically-independent configurations representative of the liquid.
	\item Calculation of phonons and more specific superconductivity properties (of the liquid).
\end{enumerate}
These steps are described in detail in this section. 

\subsection{Electronic-structure calculations}

% NOTE: I am not certain about the smearing with (for Brillouin-zone integrations), though I *suspect* it was $0.02$ Ry (standard setting).
%
All calculations were performed using the {\qe} ({\qeabbrev}) density-functional theory \cite{RevModPhys.87.897} (DFT) code \cite{0953-8984-21-39-395502}. See Ref.\ \onlinecite{PhysRevB.102.224108} (and its Supplementary Information) for a discussion on the justification of the use of DFT, and some of the following settings, to study atomic hydrogen. The pseudopotential approximation based on the projector augmented wave method \cite{PhysRevB.59.1758} was used \cite{pseudopotential_H_PBE_PAW} to replace the bare Coulomb potential of the protons. The Perdew--Burke--Ernzerhof generalized gradient approximation exchange--correlation functional \cite{PhysRevLett.77.3865} was used. {\qeabbrev} is based on a plane-wave basis set. Kinetic-energy cutoffs for the wavefunction and charge density and potential are specified in the context of specific calculations below. The same is done for Brillouin-zone sampling ({\kvec} points). The smearing scheme of Methfessel--Paxton \cite{PhysRevB.40.3616} was used for integrations based on this sampling. Note that this is ``cold'' (zero-temperature) smearing technique, but this does not affect the results in this work since finite-temperature free energies are not needed. 

\subsection{Geometry optimizations}  

Pressures were chosen over which hydrogen is expected to first be atomic, metallic, and a liquid. (Ground-state) pressures were chosen every $100$ GPa from $350$ to $850$ GPa. 
% phase diagram: A dashed line is shown in the phase diagram (at $500$ K --- to be discussed below) in Fig.\ \ref{fig:phase_diag} showing this range.

% NOTE: 16 atom cells.
%
% NOTE: ``one of the most dynamically unstable structures'' comes from Fig. S5 in the PRL (SI). 
%
The starting (ground-state) structures are unit cells of the body-centered cubic lattice. This lattice was chosen, as it has been found \cite{PhysRevLett.106.165302} to be one of the most dynamically unstable structures over the pressure range considered. Therefore, it should therefore immediately melt to liquid. Indeed, earlier simulations of the melting line of atomic hydrogen \cite{Xu_1994, PhysRevLett.74.626, PhysRevE.54.768} find that this lattice may melt at surprisingly low temperatures over the considered pressure range. 

The stationary point of each structure (both lattice vectors and ion positions) at each pressure was found by performing constant-pressure geometry \emph{optimizations}. Recommended kinetic-energy cutoffs of $46$ Ry for the wavefunction and $221$ Ry for the charge density and potential were used for the chosen pseudopotential. For Brillouin-zone sampling, at least $5{\times}5{\times}5$ (shifted) {\kvec} points (for a $16$-particle system, for example; this number is scaled precisely for larger ones) were used. Optimizations were done using the Broyden--Fletcher--Goldfarb--Shanno algorithm \cite{Fletcher:1987:PMO:39857}, as implemented within {\qeabbrev}. Pressures were converged to within $0.1$ GPa.

\subsection{Molecular dynamics}

% NOTE: Melting temperatures:
%
% doi:10.1021/jp403885h : ~350 K, with a slight positive slope
% Chen2013              : ~300 K, and flat
% PhysRevB.92.104103    : 350 K, and flat beyond 500 GPa
%
% NOTE: The importance (temperature) of nuclear quantum effects was estimated by comparing classical to quantum melting temperatures.
%
Atomic configurations representative of liquid atomic hydrogen were generated consistent with the canonical ensemble (constant $NVT$ where $N$, $V$, and $T$ are number of particles, volume, and temperature, respectively). Several simulations were carried out using a number of particles from $16$ to $250$. Comparisons between simulations were made as a test of convergence with respect to system size. Initial volumes were determined from the geometry optimizations (discussed above). Choosing the temperature involves several considerations: Relevant liquid configurations are those at or below the superconducting critical temperature, but this is \textit{a priori} unknown. Therefore, in order to determine whether this even occurs at all, temperatures should be chosen just above melting. Classical melting temperatures are calculated \cite{doi:10.1021/jp403885h, PhysRevB.92.104103} around $350$ K and flat or slight positive slope with pressure over the range considered herein. In addition, the classical superheating degree is calculated \cite{PhysRevB.92.104103} to be about $100$ K. With all of these considerations, a temperature of $500$ K was chosen. Note that these calculations also show an importance of nuclear quantum effects of just over $100$ K. This suggests that the $500$ K (classical) liquid may be representative of the actual (quantum effects included) at lower temperatures. Temperature was controlled using the Andersen thermostat \cite{doi:10.1063/1.439486}, a thermostat which is consistent with the canonical ensemble. This thermostat is specified by a single parameter --- the collision frequency $\nu$, which was set to a relatively low value (to be discussed below) of $0.002$ Ry.

% NOTE: The number of k points is justified based on the appropriate scaling from our atomic melting calculations: a (4 x 4 x 6) (250 atoms) unit cell was used for these with a $2^3$ shifted k points \cite{McMahon_H_melting_2017}.
%
Configurations were generated from Born--Oppenheimer molecular dynamics simulations. These were based on DFT. Kinetic-energy cutoffs and Brillouin-zone sampling were the same as for geometry optimizations (discussed above). The time step was set to as $10$ Ry ($1$ a.u.\ $=$ $4.8378 \times 10^{-17}$ s). Simulations were carried out in two steps: a first one of $2048$ steps to equilibrate (melt to a liquid), followed by a second one of $16384$ steps to generate enough liquid configurations for statistical analysis (discussed below).
%
% NOTE: When specifying k-points (below), the assumption ``for the $16$-particle system'' remains.

\subsection{Liquid configurations}
%\label{sec:methods:liq_conf}

% NOTE: For the calculations (``averages'' --- discussed below), five configuations were selected (and used). 
%
From the generated configurations (discussed above), random ones were selected as representative of the liquid. Properties of the liquid were then calculated as averages over these, with the mean and sample standard deviation of the mean reported. While this is correct for many properties of interest, careful consideration must be made for dynamical ones (as static configurations are only approximate, if at all). These are discussed in context below.

% NOTE: Often local minima are undesirable, but in this case they are wanted.
%
% NOTE: I *suspect* that the k points are unshifted (which would be necessary for the following electron--phonon calculations). 
%
Note that at finite temperature, a configuration of a (classical) liquid can be considered as a random thermal fluctuation about any inherent structure. Because it is this local potential-energy minimum that is of interest in the calculation of properties, less-aggressive \emph{relaxations} (in this case, at constant volume) of these configurations were performed. These were carried out using the damped Verlet algorithm \cite{PhysRev.159.98}, as implemented within {\qeabbrev}. Forward looking to the following calculations, settings for the DFT calculations were increased for convergence. Cutoffs were increased to $57.5$ and $345.5$ Ry for the wavefunction and charge density and potential, respectively. The number of {\kvec} points for Brillouin-zone sampling was increased to $16{\times}16{\times}16$ (unshifted), which is sufficiently dense to include both the full (fine) grid and an $8{\times}8{\times}8$ (course) one, which are both needed (as discussed below).
%
% NOTE: The above settings were used for all following calculations, unless otherwise specified. (This is implied by the statement ``Forward looking to the following calculations''.)

\subsection{Phonons}

% NOTE: This comparison is to justify a fundamental assumption in this work is that the solid configurations act as a stand in for the actual liquid and give representative electron-electron phonon interactions for the liquid. This is because the configurations are treated as a solid to be able to calculate the electron-phonon properties.
%
Phonons are one of the main properties of interest. Of particular interest is the phonon density of states $F(\omega)$ as a function of frequency $\omega$. These were calculated by two approaches.

Phonons of the dynamical liquid were calculated directly from the molecular dynamics simulations, by a Fourier transform of the velocity autocorrelation function. Note that the Andersen thermostat is stochastic, and so comparisons were made against the Berendsen thermostat \cite{doi:10.1063/1.448118} (a thermostat which is deterministic, but not rigorously consistent with the canonical ensemble), to ensure that the value of $\nu$ is low enough such that it does not significantly affect correlations. 

% NOTE: Convergence tests with denser k-point sampling shows that an (8 x 8 x 8) grid is converged (for calculating phonons).
%
% NOTE: Convergence tests at (8 x 8 x 8) q-points show that (4 x 4 x 4) is converged.
%
% NOTE: Minimum and maximum phonon energies and the step size are unknown.
%
Due to the small proton mass, the magnitude of the phonon frequencies is expected to be considerable in hydrogen. On the timescale of translation (by diffusion through the liquid), the atoms will have oscillated a number of times. This suggests that static configurations can be used as representative of the liquid. This was verified by comparing the following calculations to the corresponding dynamical ones (discussed above).
% SI: An example showing both of these is given in Supplementary Fig.\ 2.
Phonons for static configurations were calculated using density-functional perturbation theory \cite{RevModPhys.73.515}, as implemented within {\qeabbrev}. These were calculated with a {\kvec}-point grid of $8{\times}8{\times}8$. A $4{\times}4{\times}4$ grid of {\qvec} points was used, to calculate the phonon density of states $F(\omega)$. This value was found to be sufficient for convergence, consistent with Ref.\ \onlinecite{PhysRevB.84.144515, PhysRevB.85.219902}. Note that these calculations are based on the harmonic approximation. Anharmonic effects in atomic hydrogen have been studied in Ref.\ \onlinecite{PhysRevB.93.174308}. As far as the specific properties of interest herein, such effects are found to have only a relatively small impact. The results should therefore be considered (in this context) at least semi-quantitative. 

\subsection{Superconductivity}
\label{sec:methods:superconductivity}

% NOTE: Very specific settings (not significantly-important for the discussion):
%
% - A Matsubara frequency cutoff at $200$ was used (an integer cutoff for the number of terms to use in the equations).
% - An increase in the Matsubara cutoff resulted in very little increase in precision (to the 100--1000th decimal place) and resulted in much longer computation time.
%
% - The phonon cutoff frequency, $\omega_c$, (in units of $\omega_{max}$) was chosen to be 3.
%
% NOTE: The gap function is denoted by $\Delta$.
%
To calculate properties associated with superconductivity, the Eliashberg equations \cite{Eliashberg_JETP-1960} were used. This theory accurately describes strong electron--phonon interactions. The important microscopic parameters that enter into this theory are the (Eliashberg) spectral function $\alpha^2 F(\omega)$ and Coulomb pseudopotential $\mu^*$. Within BCS theory, the electron--phonon interaction is the source of attraction that binds two electrons into a paired state. $\alpha^2 F(\omega)$ describes this coupling, $\alpha^2(\omega)$ (it is squared, since two electrons are coupled) and $F(\omega)$ (discussed above). $\mu^*$ describes the strength of the effective Coulomb repulsion between the coupled electrons. The equations were numerical solved \cite{numerical_Eliashberg_2006} for the gap function, to find the superconducting critical temperature $T_c$ (the temperature at which this function goes to zero). 

% NOTE: Convergence of $\alpha^2 F(\omega)$ was done by eye, to when the coupling constant converged to a value or had a minimum.
%
% NOTE: the density of states at the Fermi level $\rho(0)$
%
$\alpha^2 F(\omega)$ was calculated using density-functional perturbation theory, as implemented within {\qeabbrev} (the only input necessary from {\qeabbrev} for this). A coarse grid of $8{\times}8{\times}8$ (unshifted) and dense one of $16{\times}16{\times}16$ (unshifted) of {\kvec} points were used for the DFT calculations. Note that these values are fully consistent with the converged geometry relaxations (dense grid), and also the phonon calculations (which are calculated in both cases on the course grid). The {\qvec}-point grid as used for the phonon calculations was also used for these, which should be \cite{PhysRevB.84.144515, PhysRevB.85.219902} sufficiently dense to converge the electron--phonon coupling. (Two {\kvec}-point grids are needed, as mentioned above, as the dense one must include all {\kvec}$ + ${\qvec} points.) $\alpha^2 F(\omega)$ was then calculated using a Gaussian broadening that led to convergence by inspection. Note that for calculations based on this function, a small contribution of (expected) imaginary frequencies were removed from the static calculations, since they are not present in the dynamical (phonon) ones. The value of $\mu^*$ as calculated \cite{PhysRevLett.78.118} from first principles for high-density atomic hydrogen is $0.089$. With an error of about $3$\% though, the standard value (for a high-density system) of $0.1$ remains reasonable, and this value is used herein. The only remaining quantity to calculate is therefore $\alpha^2 F(\omega)$. Finally, the density of states at the Fermi level has been calculated \cite{YAN20111264} for atomic hydrogen over the pressure range considered. The value does not change much with pressure, and a value of $0.45$ states/Ry/spin was chosen.

% NOTE: The critical temperature was also calculated for solid Cs-IV, to compare with the liquid to test if they do have similar properties at the same pressures.
%
\section{Results}
\label{sec:results}

(Finite-temperature) pressures were calculated from the molecular dynamics simulations. Those for each of the simulations is shown in Table \ref{tbl:P}.
\begin{table}
    \centering       
    \begin{tabular}{|c|c|c|c|c|c|}
        \hline
        $386.8(4)$ & $485.0(4)$ & $584.1(4)$ & $683.9(4)$ & $783.7(4)$ & $883.7(3)$ \\ 
        \hline 
    \end{tabular}     
    \caption{Pressure values (in GPa) (at $500$ K) considered in this work.}
    \label{tbl:P}
\end{table}
The range over the calculations is $386.8(4)$--$883.7(3)$ GPa. Note that while these pressures (technically) correspond to $500$ K, they are considered \emph{the} pressure values for the purpose of plots (e.g., as a function of pressure), discussion, etc. In addition, the uncertainties are generally smaller than the symbols used in these plots, and are thus omitted thereon. 

\subsection{Phonons}
\label{sec:phonon}

\begin{figure}[h]
    \centering
    \includegraphics[width = 7.5cm]{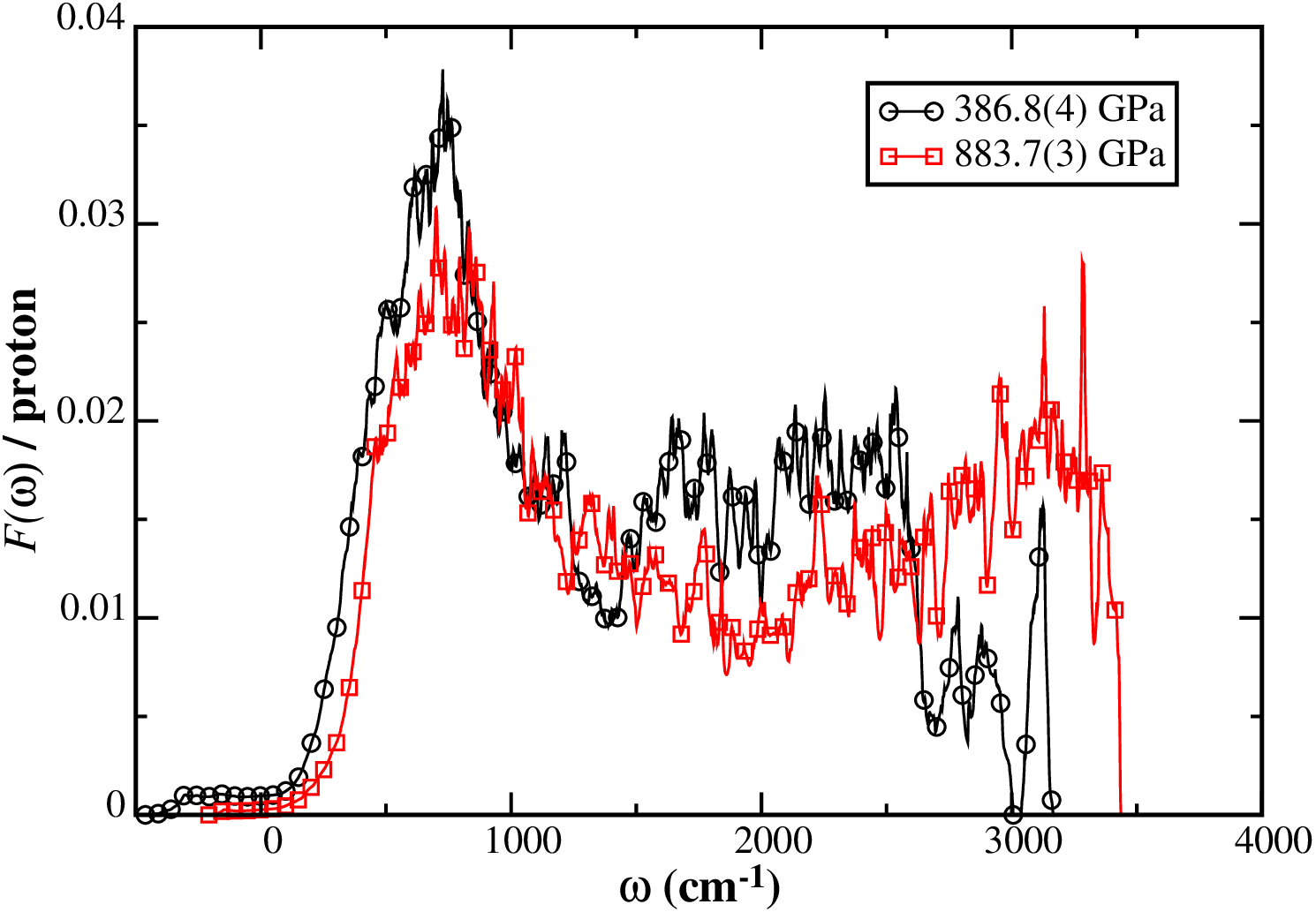}
    \caption{Phonon density of states $F(\omega)$ as a function of frequency $\omega$ at pressures of $386.8(4)$ and $883.7(3)$ GPa. Shown are those for a (random) single configuration for each pressure. Note that $F(\omega)$ has been normalized per proton, for quantitative comparisons here and below.} 
    \label{fig:phdos}
\end{figure}

% NOTE: The "similarity of phonon spectra" can be seen in CMT's thesis Fig. 6.5.
%
$F(\omega)$ of (random, as examples for discussion purpose) single configurations from simulations of liquid metallic hydrogen at $386.8(4)$ and $883.7(3)$ GPa (the end points of pressure considered) are shown in Fig.\ \ref{fig:phdos}.
Note that, at a given pressure, while the atomic configurations appear qualitatively different (by inspection), the $F(\omega)$ are similar. This can be seen, for example, by comparing Fig.\ \ref{fig:phdos} with the following comparison between the liquid and solid phases [which plots a different $F(\omega)$ for the liquid for this comparison]. Any random configuration may therefore be considered representative of the liquid, for the purpose of discussion. 

The magnitude of phonon frequencies become extremely large, reaching up to a few thousand cm$^{-1}$. Even $\langle \omega \rangle$ is large, around $1000$ cm$^{-1}$. This is considerable, considering for conventional metals $\langle \omega \rangle \approx 70$ cm$^{-1}$. This is expected though, considering the small mass of the proton that should result in high-frequency phonons. Calculations \cite{PhysRevLett.106.165302} for the solid atomic phase have shown this as the basis for large zero-point energies, for example.

% NOTE: For example, around $684$ GPa, the phonon spectra of the liquid agrees well with solid Cs-IV, in particular the highest frequency peak matches up the best.
%
$F(\omega)$ also displays significant structure. In particular, there is considerable phonon density at both low and high frequencies. This suggests that the liquid supports both strong acoustic and optical phonons, respectively. 

As the pressure is increases, the high-frequency phonons shift higher, which results in an even larger separation of the low- and high-frequency modes. The greater compression probably leads to deeper and more narrow potential-energy wells locally, which leads to larger vibrational frequencies. This effect should be less significant at low frequencies, for which the vibrational physics is different (atoms moving in phase).

\begin{figure}[h]
    \centering
    \subfloat[][$386.8(4)$ GPa]{\includegraphics[width=0.45\textwidth]{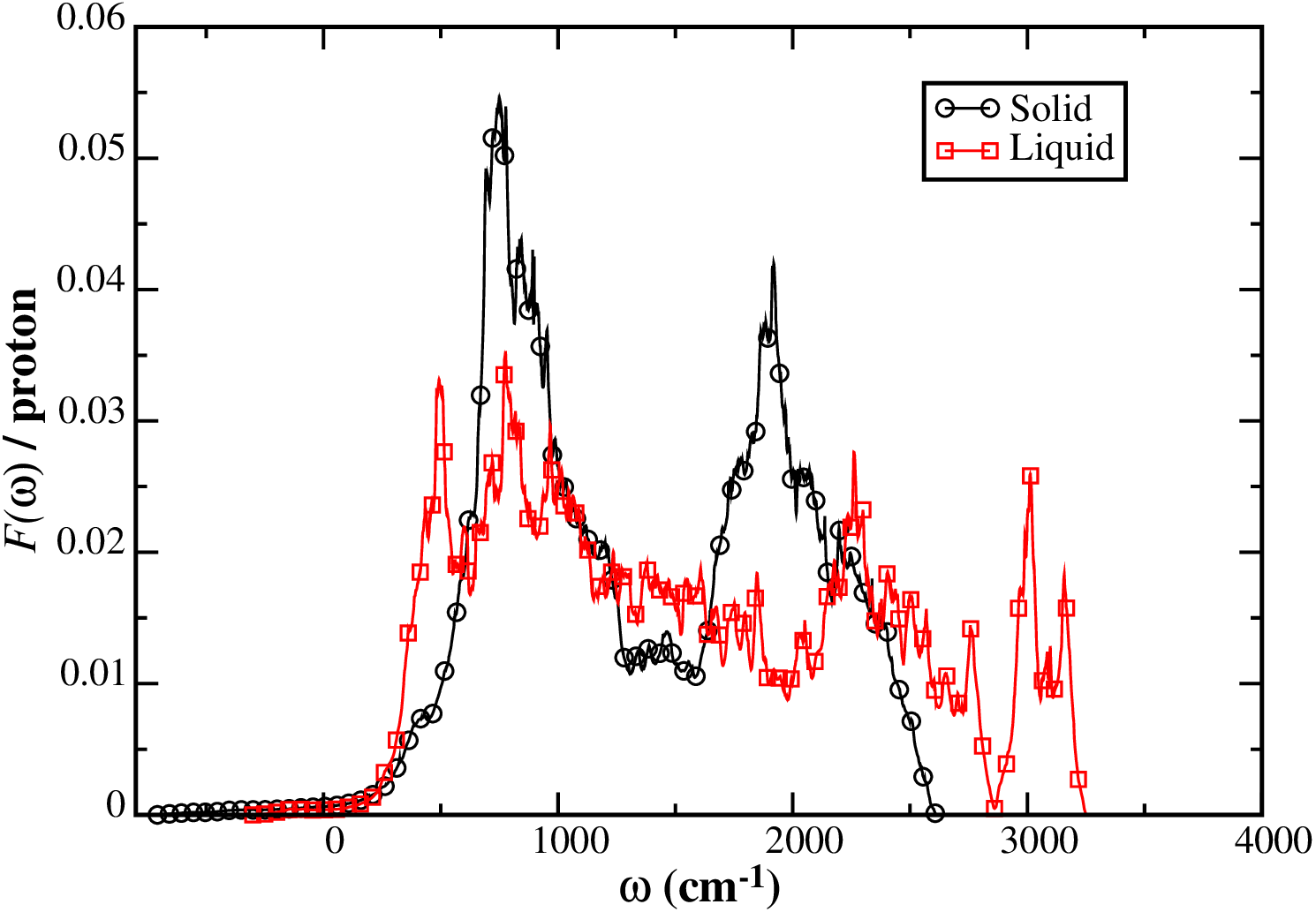}}
    \subfloat[][$883.7(3)$ GPa]{\includegraphics[width=0.45\textwidth]{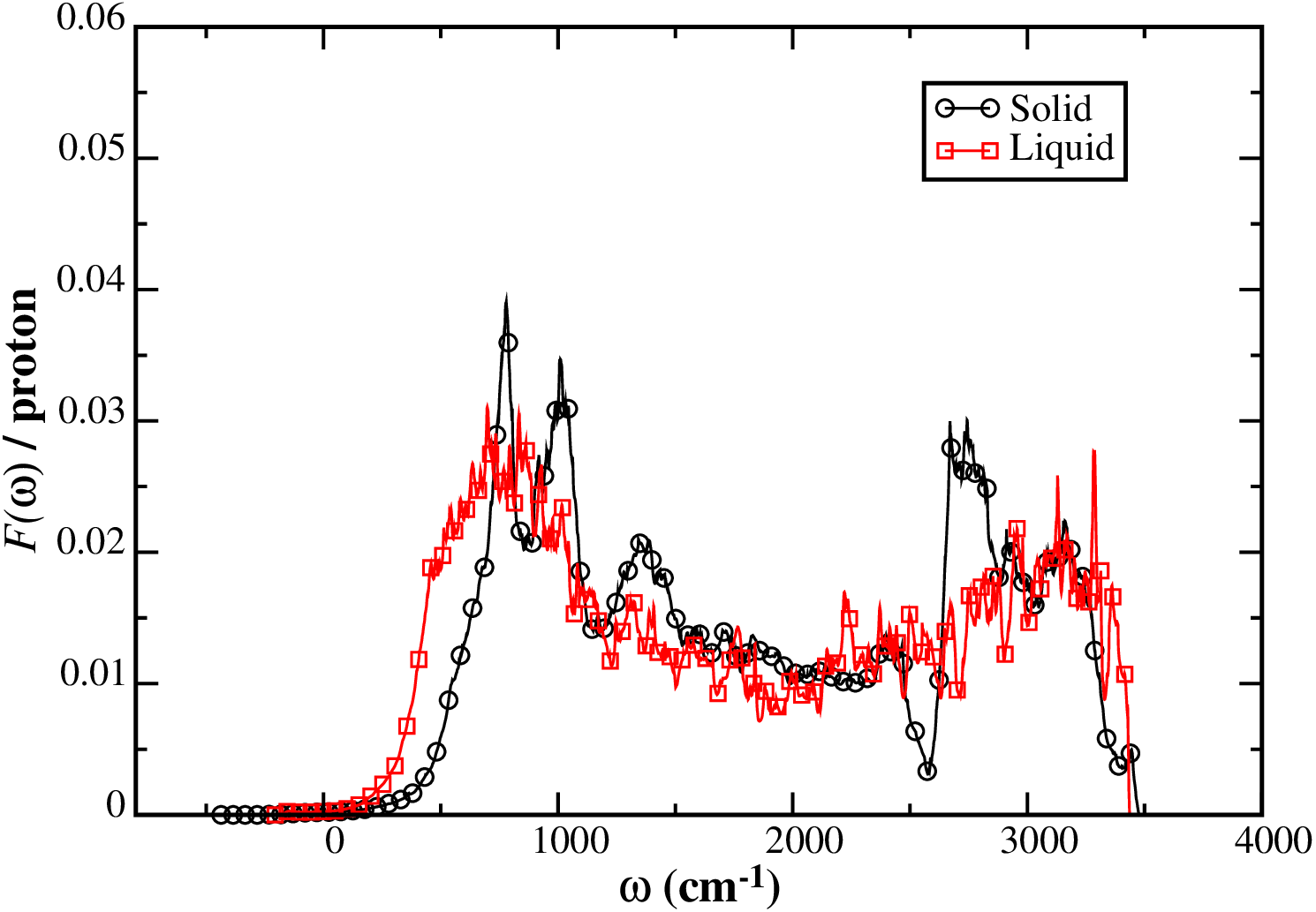}}
    \caption{Comparison of $F(\omega)$ between the solid and liquid phases.} 
    \label{fig:Fw_solid_liquid}
\end{figure}

A comparison of the liquid phonons to those calculated for the solid phase \cite{PhysRevB.84.144515, PhysRevB.85.219902} shows that they are qualitatively similar. Recalculation of also the solid data at the considered pressures and calculation settings to provide the same level of convergence is shown in Fig.\ \ref{fig:Fw_solid_liquid}.
The agreement becomes noticeably better at higher pressures. At lower pressures, the high-frequency phonons in the solid occur at lower frequencies. At the higher pressures, the agreement also seems to be better at high frequencies. This is probably the result of that the low frequencies, acoustic phonons are those in which the ions move in phase, which in a liquid are expected to be much less likely. Overall, the agreement is consistent with the suggestions of Jaffe and Ashcroft \cite{PhysRevB.23.6176}.

\subsection{Electron--phonon coupling}
\label{sec:elph}

In hydrogen, the electron--ion interaction is expected to be significant whether it is in the solid or liquid (or other) phase, as it arises from the bare Coulomb interaction. The electron--phonon interaction is therefore also expected to be high. $\alpha^2 F(\omega)$ can be used to quantify this.

\begin{figure}
    \centering
    \includegraphics[width = 7.5cm]{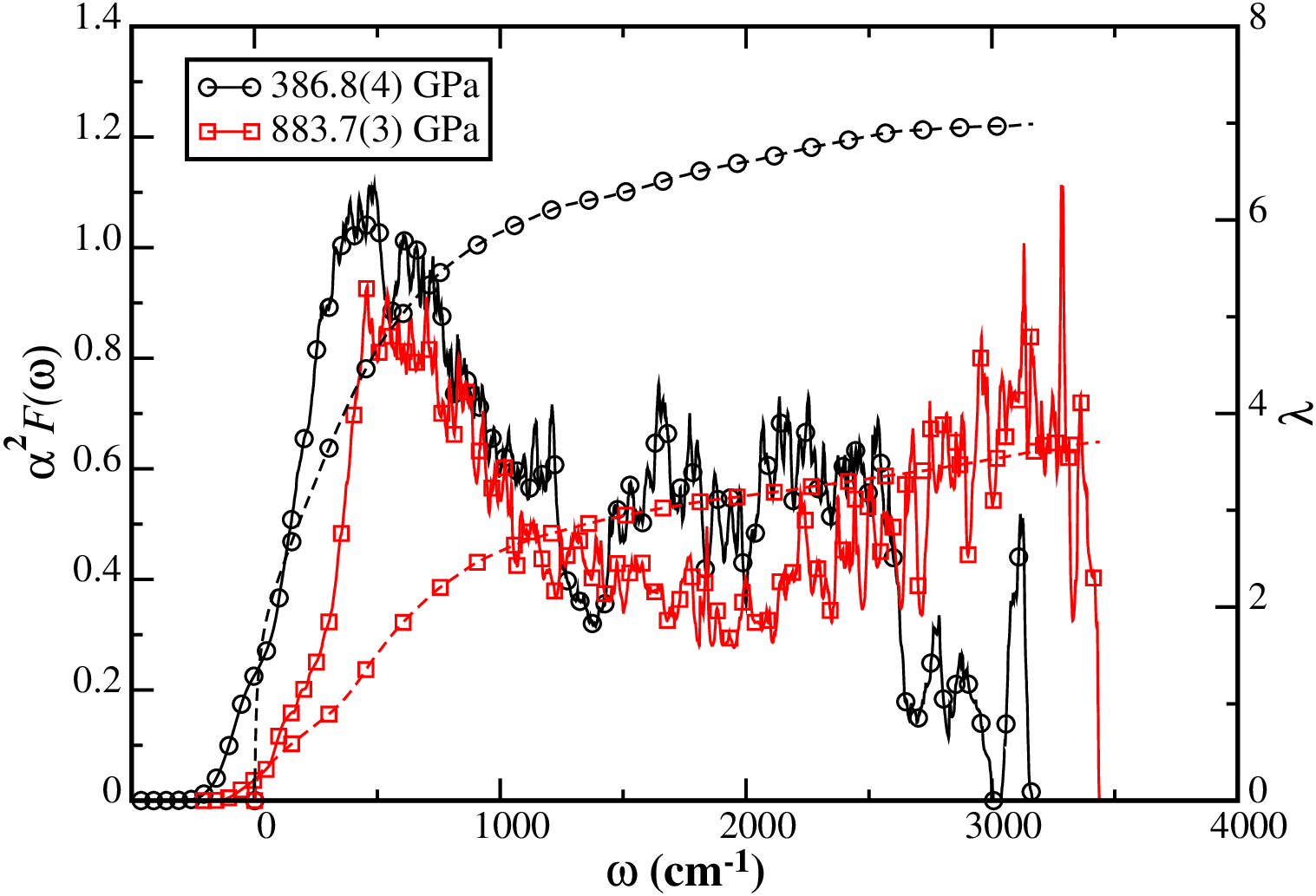}        
    \caption{Eliashberg spectral function $\alpha^2 F(\omega)$. The cumulative value of the expression in Eq.\ (\ref{eq:lambda}) in the text is plotted using dashed lines. The result of this integral over all $\omega$ is denoted by the electron--phonon coupling constant $\lambda$.} 
    \label{fig:a2f} 
\end{figure}

% NOTE: A comment left by CMT suggests that negative frequencies are included in lambda in the following plot.
%
Figure \ref{fig:a2f} shows $\alpha^2 F(\omega)$ for the two pressures, $386.8(4)$ and $883.7(3)$ GPa, considered above.
$\alpha^2 F(\omega)$ follows the same trend as the phonon spectra in Fig.\ \ref{fig:phdos}. This can be understood by considering that if $\alpha^2(\omega)$ is relatively flat (which means that the electron--ion interaction is relatively frequency-independent, and which appears to be the case here), then it is similar to $F(\omega)$.

While $\alpha^2 F(\omega)$ contains all of the relevant information, consider the cumulative integral
\begin{equation} 
\label{eq:lambda}
    2 \int_{\omega_\text{min}}^{\omega_\text{max}} d\omega ~ \frac{\alpha^2 F(\omega)}{\omega}
\end{equation}
where ${\omega_\text{min}}$ and ${\omega_\text{max}}$ are minimum and maximum cutoff frequencies, respectively. This provides qualitative insight into the electron--phonon coupling over $[{\omega_\text{min}}, {\omega_\text{max}}]$. This integral is plotted on the right axis in Fig.\ \ref{fig:a2f} from $0$ to $\omega$. At low frequencies, there is a rapid rise in it. It then only steadily does so at higher ones. It can therefore be concluded that lower frequencies are actually contributing the most to this coupling.

\begin{figure}[h]
    % \captionsetup[subfigure]{labelformat=empty}
    \centering
    \includegraphics[width = 7.5cm]{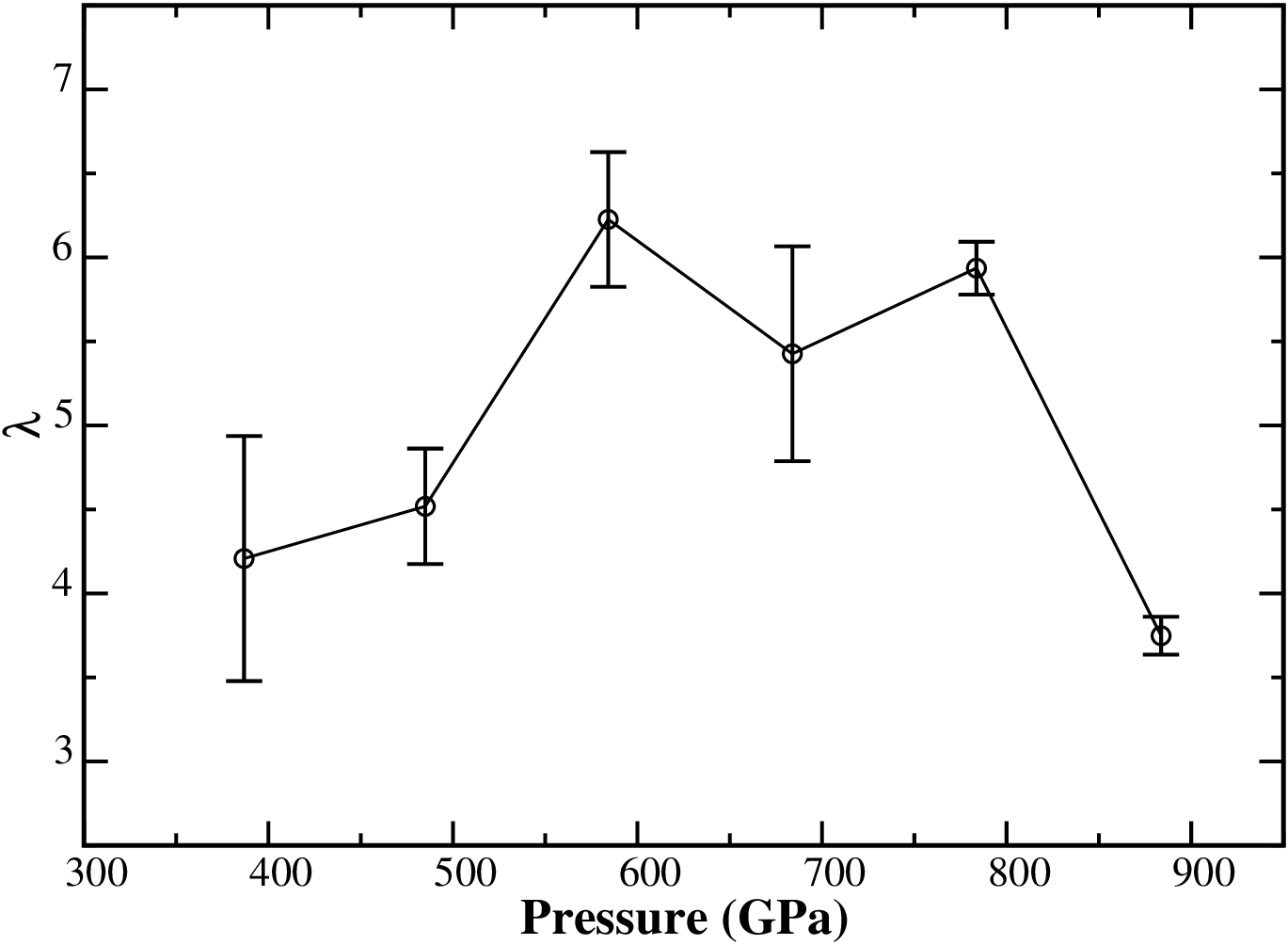}
    \caption{The value of $\lambda$ as a function of pressure.}
    \label{fig:lambda}
\end{figure}

% NOTE: This is excluding imaginary frequencies.
%
% NOTE: Comparing the values of $\lambda$ from Fig.\ \ref{fig:a2f_press_a} and Fig.\ \ref{fig:lambda} at $386.8$ GPa, the final $\lambda$ contribution is much higher then the value reported in Fig.\ \ref{fig:lambda}. This is due to the average that was taken. The particular configuration chosen in Fig.\ \ref{fig:a2f_press_a} happens to have the highest $\lambda$ value of the five configurations. So the average value in Fig.\ \ref{fig:lambda} is lower.
%
% NOTE: References for the sentences describing coupling regimes were moved to the end and outside (of them), so that they do not get mistaken for exponents.  
%
The total electron--phonon coupling constant is given by the range $[0, \infty)$, which is denoted by $\lambda$. That is, the total cumulative result of Eq.\ (\ref{eq:lambda}), which is also shown and indicated in Fig.\ \ref{fig:a2f} (for the two considered configurations). Figure \ref{fig:lambda} shows the average values of $\lambda$ over the pressure range considered.
The range of $\lambda$ is about $4$--$6$, which is very high. For common metals, $\lambda$ is below $0.5$. \cite{PhysRev.125.1263} The high value of $\lambda$ puts it above the strong-coupling regime, given as $\lambda > 1$. \cite{PhysRev.167.331} There is an increase of $\lambda$ with pressure, resulting in a maximum around $600$--$800$ GPa. 

The magnitude of $\lambda$ is significantly higher (by a factor of about $2$--$3\times$) in the liquid than in the solid phase. For example, the calculation \cite{PhysRevB.93.174308} of $\lambda$ (including anharmonic effects) is $1.63$ at $500$ GPa. At least at the intermediate and higher pressures considered, this can be understood again considering $F(\omega)$, which has a greater density of phonons at low pressures (the low-frequency ``peak'' occurs at lower pressures, and the distribution is broader) (see again Fig.\ \ref{fig:Fw_solid_liquid}); this extends analogously to $\alpha^2 F(\omega)$ (as discussed above), which causes Eq.\ (\ref{eq:lambda}) to integrate larger (because of the lower $\omega$). The trend with pressure is similar between the two phases though. Calculations \cite{PhysRevB.84.144515, PhysRevB.85.219902} (not including anharmonic effects, but nonetheless agree well, at least at $500$ GPa) show an increase with pressure up to $700$ GPa before then decreasing.

\subsection{Critical temperature}
\label{sec:crit}

\begin{figure}[h]
    \centering
    \includegraphics[width = 7.5cm]{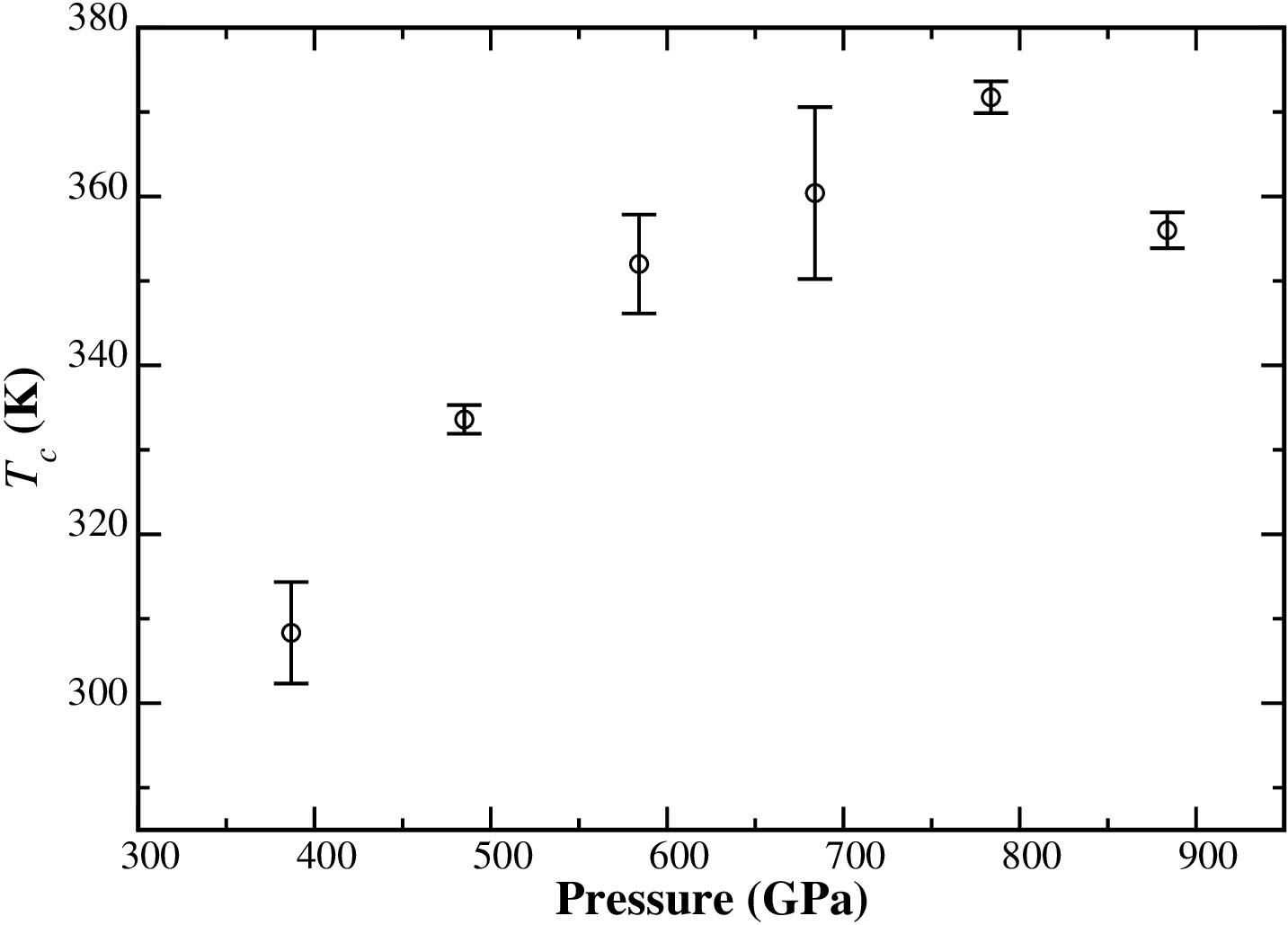}
    \caption{Superconducting critical temperatures $T_c$ as a function of pressure.} 
    \label{fig:Tc}
\end{figure}

With the above information (and additional considerations discussed in Section \ref{sec:methods:superconductivity}), the Eliashberg equations can be solved to find $T_c$. The value of $T_c$ as a function of pressure is shown in Fig.\ \ref{fig:Tc}.
The values of $T_c$ are very high. The lowest value is $308(6)$ K at $386.8(4)$ GPa. It then increases semilinearly to $372(2)$ K at $783.7(4)$ GPa, before decreasing. This trend is consistent with those for the other quantities (discussed above), and can thus be understood in the context of those. 

Again comparing to the solid phase, both the quantitative value (at $500$ GPa) \cite{PhysRevB.93.174308} and trends with pressure \cite{PhysRevB.84.144515, PhysRevB.85.219902} are very similar. The physics though is slightly different. The phonon spectrum in the solid (understandably) has considerably more structure and which occur at higher frequencies (at least at relatively lower and intermediate pressures). This alone should result in larger values of $T_c$. However, it is for this precise reason that $\lambda$ has significantly lower values. Considered together, the $T_c$ values for the liquid and solid phases end up being similar.

\section{Discussion \& Conclusions}
\label{sec:discussion}

The superconducting critical temperatures of liquid atomic hydrogen have been calculated. Over the pressure range $386.8(4)$--$783.7(4)$ GPa, $T_c$ increases from $308(6)$ to $372(2)$ K with a maximum uncertainty of $10$ K; it then decreases to $356(2)$ K at $883.7(3)$ GPa. Comparisons to the solid phase show that the critical temperature is not significantly changed between the two phases, though the physics behind their superconductivity is different.

% NOTE: 
%
% Y. Ma (2013): ~250 K at 400 GPa, with a negative slope with pressure
% Needs (2013): <200 K, with a negative slope with pressure
% (2015)      : 200--250 K, with a flat variation with pressure
%
While the values of $T_c$ are remarkably high, whether this phase is realizable depends on the melting temperature of the solid. (The possibility of a metastable liquid is not considered in the following discussion, though this could be another approach to realization.) On the lower end of pressures considered, calculations predict the existence of a quasi-molecular mC24 phase \cite{doi:10.1021/jp301596v}, in the pressure range between the stability fields of the molecular $Cmca$-$4$ phase \cite{Edwards_1996} and atomic Cs-IV one \cite{PhysRevLett.106.165302}. At higher pressures, the melting line of atomic hydrogen (including quantum effects) has been calculated \cite{doi:10.1021/jp403885h, PhysRevB.92.104103} to melt between around $200$--$250$ K and with a relatively flat change (or slight decrease) with pressure over the pressure range considered. (This decrease has also been calculated in Ref.\ \onlinecite{Chen2013}, though some of the quantitative aspects of these calculations have been shown \cite{PhysRevB.92.104103} to be incorrect.) And the melting temperature of the mC24 phase is systematically lower \cite{doi:10.1021/jp403885h}. These lines are shown (near $225$ K) in Fig.\ \ref{fig:phase_diag}. Calculations \cite{doi:10.1021/jp403885h} predict a second extremum (a minimum, in this case) in the melting line at the intersection of the melting lines of the molecular and atomic phases. Classically, this occurs at approximately $432$ GPa and $367$ K; but this is almost certainly reduced by nuclear quantum effects (that for the atomic phase alone would suggest about $100$ K), so for the purpose of discussion it will be considered near $300$ K. Note that the melting temperature (of the molecular phase) then increases with a reduction in pressure. This line (with the aforementioned suggested effects) is also shown in Fig.\ \ref{fig:phase_diag}. 

% NOTE: The following estimates of the observation of the LLPT come from Fig. 2 in the SI of the references PNAS.
%
Another consideration is that the solid melts to a metallic state. A metallic liquid is necessary for superconductivity. There is a liquid--liquid phase transition (between molecular and atomic liquids), as also shown on Fig.\ \ref{fig:phase_diag}. Calculations \cite{PhysRevB.102.195133} show closure of the fundamental electronic gap strongly correlates with the onset of molecular dissociation. This phase transition is perhaps most accurately and precisely known from calculations \cite{Pierleoni4953}, and this line is also shown in Fig.\ \ref{fig:phase_diag}. Along the $600$ K isotherm (the lowest temperature considered in the calculations), this transition is calculated to occur between approximately $267$ and $275$ GPa. Considering the decrease in temperature with pressure, an extrapolation to higher pressures would put it in the region of the aforementioned second extremum in the hydrogen melting line. 

With all of the above considerations taken into account, it can be concluded that liquid atomic hydrogen will exist in a superconducting state. The lowest pressure at which this should be observable is near the second extremum at approximately $432$ GPa and around $300$ K. It is in this region where solid atomic hydrogen should melt to a liquid atomic phase. With an increase in $T_c$ with pressure (up to a maximum), and a flat or slight decrease in melting temperature, this state should also exist at higher pressures. 

The main challenge is experimental, being able to reach the necessary thermodynamic conditions. Recent developments \cite{PhysRevLett.119.075302} in heated diamond anvil cell techniques allow the study of hydrogen up to $300$ GPa at $295$--$1000$ K. This is approaching the necessary pressure and temperature conditions discussed herein. And the results of this work provide ones over which to look for this state.

% =========================
% BIBLIOGRAPHY
% =========================

%\def\urlprefix{}
%\def\url#1{}
%\def\doiprefix{}
%\def\doi#1{}
%\def\eprintprefix{}
%\def\eprint#1{}
% \bibliographystyle{naturemag}
%\bibliography{../latest.bib,../phase_diagram.bib,../phase_diagram_misc.bib}


% =========================
% ACKNOWLEDGMENTS
% =========================

\begin{acknowledgments}
We thank members of the McMahon Research Group, for help with the hydrogen phase diagram. J.\ M.\ M.\ acknowledges startup support from Washington State University and the Department of Physics and Astronomy thereat.
\end{acknowledgments}

% =========================
% CONTRIBUTIONS
% =========================

% \

% \input{contributions}

% =========================
% ADDITIONAL INFORMATION
% =========================

% \

%\input{additionalinfo}

\end{document}